\magnification = 1020
\baselineskip16pt
\abovedisplayskip 6pt plus2pt minus3pt
\belowdisplayskip 6pt plus2pt minus3pt
\raggedbottom
\vsize=9.55 true in
\hsize=6.1 in
\font\bmit=cmmib10 \textfont9=\bmit \def\bmit{\fam9 }
\font\bx=cmbsy10 \textfont10=\bx \def\bx{\fam10 }
\mathchardef\alpha="710B
\mathchardef\epsilon="710F
\mathchardef\pi="7119
\mathchardef\sigma="711B
\mathchardef\tau="711C
\mathchardef\mu="7116
\mathchardef\nabla="7272
\parskip=8pt plus3pt minus2pt
\interlinepenalty = 100
\def\boxit#1{\vbox{\hrule\hbox{\vrule\kern5pt
   \vbox{\kern5pt#1\kern5pt}\kern5pt\vrule}\hrule}}
\def\sqr#1#2{{\vcenter{\vbox{\hrule height.#2pt
   \hbox{\vrule width.#2pt height#1pt  \kern#1pt
      \vrule width.#2pt}
     \hrule height.#2pt }}}}

\def\smt{\let\rm=\sevenrm \let\bf=\sevenbf \let\it=\sevenit
    \let\sl=\sevensl \baselineskip=9.5pt minus .75pt  \rm }

\centerline {\bf Poincar\'e group operators with 4-vector position }

\rightline{Shaun N Mosley,\footnote{${}^*$} {E-mail:
shaun.mosley@ntlworld.com }
Alumni, University of Nottingham, UK }

\vskip 0.6in

\noindent {\bf Abstract }\hfil\break
\noindent We present a new set of massless Poincar\'e group
operators Hermitian with respect to the $ 1 / r \, $ inner product space,
which have quasi-plane wave energy-momentum eigenfunctions having velocity
$ c \, $ along their axis of propagation.
These are constructed by means of a unitary transformation from a known set
of massless Poincar\'e group
operators of helicity $ s = 0 , \pm {1 \over 2} , \pm 1 ... $
The position vector $ {\bf r} \, $ is the space part of a null 4-vector.

\beginsection I. Introduction

The standard relativistic quantization procedure [1] involves the
integral operator
$$ \eqalignno{
&\omega \equiv \surd ( m^2 \, + \, p^2 )
=  \surd ( m^2 \, - \, \nabla^2 ) \,  \cr
\noalign{\noindent in units with $ \hbar = c = 1 \, ,$ then }
&H \equiv p^0 = \omega \, , \qquad \qquad \qquad
{\bf p} = - \, i \, {\bx \nabla} \, & (1) \cr
\noalign{\noindent which together with the Lorentz generators of boosts
and rotations }
&{\bf K} = {1 \over 2} \, ( {\bf r} \, H \, + \, H \, {\bf r} ) \, ,
\qquad  {\bf J} = - i {\bf r}   \times {\bx \nabla} \, & (2) \cr
  }  $$
make up the ten Poincar\'e group operators.
There are a number of well-known difficulties with this procedure, of which
we mention the following.
1) The conserved inner product in configuration space is [2]
$$ \langle \phi \, | \, \psi \rangle\,
= \, \int d^3 {\bf r} \; [ \phi^* \, \omega \psi \,
+ \, ( \omega \phi^* ) \, \psi \, ] \, , \eqno  (3)  $$
but the quantity $ [ \psi^* \, \omega \psi \,
+ \, ( \omega \psi^* ) \, \psi \, ] \, $ can be negative even
for superpositions of positive energy solutions satisfying
$ i \partial_t \psi = + \, \omega \, \psi \, .$ And
2) The position operator $ {\bf r} \, $ is not Hermitian,
and must be replaced by the Newton-Wigner [3] position operator
$ {\bf r}_{NW} \, $
the eigenfunctions of which are not delta functions.

We set out in this paper an alternative quantization procedure which
allows the `natural' position operator $ {\bf r} \, .$ Our aims are
\hfil\break \noindent
(A) to find an alternative set of Poincar\'e group operators: while the
momentum operator will not have the simple form $ ( - i {\bx \nabla} ) \, ,$
we require that  \hfil\break \noindent
(B) the eigenfunctions of the new energy-momentum operators $ P^\lambda \, $
represent waves which propagate at velocity $ c \, $ along their propagation
axis, \hfil\break \noindent
(C) the Hamiltonian $ P^0 \, $ will be a positive operator,
\hfil\break \noindent
(D) the Poincar\'e group operators will be Hermitian with respect to a
inner product space $ {\cal L} \, $ such that
$ \langle \psi \, | \, \psi \rangle \, $ is the integral of a positive
definite density, and \hfil\break
\noindent
(E) the position operator $ {\bf r} \, $ must be Hermitian with respect to
$ {\cal L} \, $ and be part of a 4-vector position operator $ r^\lambda \, .$

In this paper we consider only mass zero representations
with helicity $ s = 0 , \pm {1 \over 2} , \pm 1 ... ,$ leaving the case
with mass to a following paper.
We start from a known [4] set of Poincar\'e group operators
$ ( a^\lambda , \, J^{\lambda \mu } ) \, ,$ the space part of
$ a^\lambda \, $ being closely related to the Runge-Lenz vector from the
theory of the Schrodinger hydrogen atom.
These operators are Hermitian with respect to the $ 1 / r \, $ inner product
space of (7) below, and furthermore the $ a^0 \, $ operator is positive.
As the eigenfunctions of $ a^\lambda \, $
are waves of variable velocity, we proceed in the next section
to find a unitary transformation
such that the the transformed eigenfunctions have plane wave character - in
that their velocity along their propagation axis is $ c \, .$

The operators [4]
$$ \eqalignno{
&a^0 = - \,  r \, {\nabla}^2 \, , \qquad \qquad \qquad \qquad \quad \; \;
{\bf a} =  - \, 2 \, ( \partial_r r) \, {\bx \nabla} \,
+ \, {\bf r} \, {\nabla}^2  \,  &  (4) \cr
&{\bf K} \equiv ( J^{10}, J^{20},J^{30} ) = - \, i \, r \, {\bx \nabla} \, ,
\qquad  {\bf J} \equiv ( J^{23}, J^{31},J^{12} )
= - i {\bf r}   \times {\bx \nabla}  &  (5)  \cr
\noalign{ \noindent obey the necessary Poincar\'e group commutation
properties: }
&[J^{\lambda \mu} , a^\nu ] = i \, ( \eta^{\mu \nu} a^\lambda
-  \eta^{\lambda \nu} a^\mu ) \, , \qquad \quad
[a^{\lambda} \, , \,  a^{\mu}] = 0 \, ,   & (6) \cr
&[J^{\lambda \mu} ,J^{\nu \rho} ]
= i (\eta^{\lambda \rho} J^{\mu \nu} + \eta^{\mu \nu} J^{\lambda \rho}
- \eta^{\lambda \nu} J^{\mu \rho} - \eta^{\mu \rho} J^{\lambda \nu}) \, .
          \cr
\noalign{\noindent These operators
$ a^\lambda , \, J^{\lambda \mu } \, $
are perhaps better known as components of a SO(4,2) group representation,
see for example [5]. Also }
&a^\lambda \cdot a_\lambda = 0 \, \cr
  }  $$
so that $ J^{\lambda \mu} , a^\nu \, $ of (4,5) are a massless spin-zero
set of Poincar\'e group operators. The
$ a^0 \, $ is a positive operator.
They are Hermitian with
respect to the $ 1/r \, $ inner product space $ {\cal L}_{1/r} \, $
$$ {\cal L}_{1/r} \, : \qquad \langle \phi \, | \, \psi \rangle_{1/r}
\equiv \int \, \phi ^* ({\bf r}) \, {1\over r} \, \psi ({\bf r}) \,
d^3 {\bf r}
= \int \, \phi ^* (r ,\theta , \phi ) \, \psi (r ,\theta , \phi )
\, r \, dr \, d\Omega \, \eqno (7) $$
which is Lorentz-invariant [6].
These $ J^{\lambda \mu} , a^\nu \, $ are to our knowledge the simplest
(non-trivial) set of configuration space Poincar\'e operators
containing only local (differential) operators.
The density $ \rho = {1 \over r} \psi^* \psi \, $ is conserved provided that
the $ \psi \, $ are solutions of $ i \, \partial_t \psi = a^0 \, \psi \, ,$
as then
$$ \partial_t ( {1 \over r} \psi^* \psi \, )
= {\bx \nabla} \cdot \, i \,
[ \psi^* {\bx \nabla} \psi \, - \, ( {\bx \nabla} \psi^* ) \, \psi \, ] \, .$$

Orthonormal eigenfunctions of $ a^\lambda $
with respective eigenvalues $ k^\lambda \equiv
(|{\bf k}| , \, {\bf k} ) \equiv (k , \, {\bf k} ) \, $ are
the Bessel functions ([4] formula (92)):
$$ \eqalignno{
&a^\lambda \, u_{\bf k} = k^\lambda \, u_{\bf k} \, \cr
\noalign{\noindent \baselineskip 16pt where }
&u_{\bf k} ( {\bf r} ) = {1\over 4 \pi}
J_0 \Big( [ - 2 k^\lambda r_\lambda]^{1/2} \Big) \, = \, {1\over 4 \pi}
J_0 \Big( [ 2 k r + 2 {\bf k} \cdot {\bf r}]^{1/2} \Big)   & (8)
    \cr } $$
which may be proved by direct differentiation. The quantity
$ ( k r + {\bf k} \cdot {\bf r} ) \, $ is non-negative by Schwartz's rule.
The orthogonality and completeness relations are [4,7]
$$ \eqalignno{
&\int \, {d^3 {\bf r}  \over r } \, u_{\bf k} ( {\bf r} ) \;
u_{\bf k'} ( {\bf r} ) = k \, \delta ({\bf k} - {\bf k}') \, , &  (9) \cr
&\int \, {d^3 {\bf k}  \over k } \, u_{\bf k} ( {\bf r} ) \;
u_{\bf k} ( {\bf r'} )
= r \, \delta ({\bf r} - {\bf r}') \, .  &  (10) \cr }   $$

Helicity $ s \, $ representations of the Poincar\'e group which
reduce to (4,5) when $ s = 0 \, $ were found by Derrick [8].
The operators $ a_s^0 , \, {\bf a}_{s} , \, {\bf K}_s , \, {\bf J}_s \, $
which also satisfy the the Poincar\'e group algebra (6) are
$$ \eqalignno{
&a_s^0 \, =  - \, r \, {\nabla}^2 \,
+ \, 2 \, s \, { i \over r - z } \, \partial_\phi \, + \, 2 \, s^2 \,
{ 1 \over r - z } \, ,  &  (11) \cr
&{\bf a}_{s} =  - \, 2 \, ( \partial_r r) \, {\bx \nabla} \,
+ \, {\bf r}  {\nabla}^2 \,
- \, s \, 4 \, i \, ( {\bf W} \times {\bx \nabla} ) \,
- \, s^2 \, 2 \, {\bmit \epsilon}_3 \, { 1 \over r - z }  \,  &  (12)  \cr
&{\bf K}_s = - \, i \, r \, {\bx \nabla} \,
- \, s \, ( {\bf \hat r} \times {\bf W} ) \, , \qquad \quad
{\bf J}_s = - i {\bf r}   \times {\bx \nabla} \,
+ \, s \, {\bf W} \, &  (13) \cr
\noalign{ \noindent where  }
&{\bf W}
= ( { x \over r - z } , \, { y \over r - z } , \, - 1 )   \, , \qquad
\partial_\phi = ( {\bf r} \times {\bx \nabla} )_3 \;   & (14) \cr
  }  $$
with $ s = 0 , \pm {1 \over 2} , \pm 1 ...$
The eigenfunctions of $ a_s^\lambda \, $ are [8]
$$ \eqalignno{
u_{s{\bf k}} ( {\bf r} ) &= {1\over 4 \pi} \,
e^{2 \, i \, s \, f ( {\bf \hat r} , {\bf \hat k} ) }
J_{2 s} \Big( [ 2 k r + 2 {\bf k} \cdot {\bf r}]^{1/2} \Big) \, .  & (15)
    \cr } $$
We give the angular phase factor
$ e^{2 \, i \, s \, f ( {\bf \hat r} , {\bf \hat k} ) } \, $
for reference in the appendix. And
$ e^{2 \, i \, s \, f ( {\bf \hat r} , {\bf \hat k} ) }
\rightarrow \, e^{2 \, i \, s \, \phi } \, $
when $ {\bf k} = ( 0 , 0 , \pm k ) \, ,$ where
$ e^{ i \, \phi } \, \equiv ( {\hat r}_1 + i {\hat r}_1 )
/ ( {\hat r}_1^2 + {\hat r}_2^2 )^{1/2} \, .$ Then the eigenfunction (15) is
$$ u_{s(0,0,\pm k)} ( {\bf r} )
= {1\over 4 \pi} \, e^{2 \, i \, s \, \phi } \,
J_{2 s} \Big( [ 2 k r \pm 2 k z ]^{1/2} \Big)  \, .
  \eqno (16) $$
The eigenfunctions $ u_{s{\bf k}} \, $ have the same orthogonality and
completeness relations (9,10) as their helicity zero counterparts.

\vskip 0.2in

\noindent{\bf 2. The unitary operator $ {\cal V} \, $ \hfil\break }
\noindent The eigenfunctions $ u_{s{\bf k}} \, $
are essentially waves in parabolic coordinates.
As massless waves proceed at
the velocity of light $ c \, $ (which is unity in our units),
we look for a unitary transformation of the
operators $ a^\lambda , \, J^{\lambda \mu } \, $
defining new operators
$ P_s^\lambda \equiv {\cal V} \, a_s^\lambda {\cal V}^{- 1} \, , \;
\overline J_s^{\lambda \mu } \equiv {\cal V} \,
J_s^{\lambda \mu } {\cal V}^{- 1} \, ,$
such that the resulting eigenfunctions $ {\cal V} \, u_{s{\bf k}} \, $ of
the $ P_s^\lambda \, $ energy-momentum operators
represent waves of velocity $ c \, $ along their
propagation axis. 

Consider
$ \psi ( u \, {\bf r'}) \, $ with $ ( - \infty < u < \infty ) \, ,$
which is the wavefunction along the line through the origin which includes
the point $ {\bf r'} \, .$
The unitary operators $ {\cal V} \, $ of (21) below map
$ \psi ( u \, {\bf r'}) \, $ onto itself, which is clearly seen
by inspection of (21d). In the appendix we show how $ {\cal V} \, $ can be
constructed from the parity $ {\cal P} \, ,$ inversion $ {\cal N} \, ,$
and Fourier transform operators $ {\cal F} \, :$
$$  \eqalignno{
&{\cal P} \, f(r , \theta , \phi )
\equiv  f (r , \pi - \theta , \phi + \pi ) \, &  (17)  \cr
&{\cal N} \, f(r , \theta , \phi ) \equiv { 1 \over r^2 } \,
f ({ 1 \over r } \,  , \theta , \phi ) \, &  (18)  \cr
&{\cal F}_{c \atop s } \, f(r , \theta , \phi )
\equiv \, \sqrt{2 \over \pi } \, \int_0^\infty \,
{\cos (  r \, t ) \atop \sin (  r \, t ) }  \;
f (t , \theta , \phi ) \, dt  \, &  (19) \cr
&{\cal F}_{{}_\pm} \,
\equiv ( {\cal F}_c \, \pm \, i \, {\cal F}_s ) \, .  \cr
   }   $$
The operators
$ {\cal V} \, , \; {\cal V}^{-1} \, $ are
$$ \eqalignno{
{\cal V}^{-1} \, g ( r , \theta , \phi ) \,
&\equiv {1 \over 2 } \, \big[
( {1 \over \sqrt{r}} {\cal F}_+ \sqrt{r} ) \,
- \, ( {1 \over \sqrt{r}} {\cal F}_- \sqrt{r} ) \,
{\cal P} \, \big] \, {\cal N} \, g ( r , \theta , \phi ) \, &  (20a)  \cr
{\cal V}  \, f ( r , \theta , \phi ) \,
&\equiv {1 \over 2 } \,{\cal N} \, \big[
( {1 \over \sqrt{r}} {\cal F}_- \sqrt{r} ) \,
- \, ( {1 \over \sqrt{r}} {\cal F}_+ \sqrt{r} ) \,
{\cal P} \, \big] \, f ( r , \theta , \phi ) \,  & (21a)  \cr
&= {1 \over \sqrt{ 2 \pi } } \, {1 \over \sqrt{r}^3 }
\int_0^\infty \, dt  \; \big[ e^{ - i \, t / r } \; \sqrt{t} \;
f ( t , \theta , \phi ) \, - \, e^{ i \, t / r } \; \sqrt{t} \;
f ( t , \pi - \theta , \phi + \pi ) \, \big]  & (21b) \cr
\noalign{\noindent \baselineskip16pt which can be written }
{\cal V}  \, f ( {\bf r} ) \,
&= {1 \over \sqrt{ 2 \pi } } \,
\int_0^\infty \, du \; \big[ \, e^{ - i \, u} \; \sqrt{u} \;
f ( u \, {\bf r}) \, - \, \, e^{ i \, u} \; \sqrt{u} \;
f ( - \, u \, {\bf r}) \, \big] & (21c)  \cr
&= {1 \over \sqrt{ 2 \pi } } \,
\int_{ - \infty }^\infty \, du \; \, e^{ - i \, u} \; \sqrt{|u|} \;
\hbox{sgn} (u) \;  f ( u \, {\bf r}) \, .  & (21d) \cr
\noalign{\noindent \baselineskip16pt
The kernel
$ \big\{ e^{ - i \, u} \; \sqrt{|u|} \; \hbox{sgn} (u) \, \big\} $
and its derivatives are continuous at  $ u = 0 \, .$
We see that
$ ({\cal V} \, f) ({\bf r'}) \, $ is essentially a Fourier
transform of $ \sqrt{r} \, f ({\bf r}) \, $ along the
line $ {\bf r} = u \, {\bf r'} , \; ( - \infty < u < \infty ) \,  .$
The $ {\cal V}^{- 1} \, $ operator can be simplified to
(corresponding to the form (21d) for  $ {\cal V} \, )$ }
{\cal V}^{-1} \, g ({\bf r})
&={1 \over \sqrt{ 2 \pi } } \,
\int_{- \infty}^\infty \, du  \;  \; { e^{ i / u }  \over \sqrt{|u|} } \;
\hbox{sgn} (u) \;  \; g ( u \, {\bf r}) \, & (20b) \cr
   }   $$

If $  \sqrt{r} \, f (r) \, $ does not converge at infinity,
the integral of (21) does not exist.
For this case the definitions of $ {\cal F}_{{}_\pm} \, $ of (19) must be
extended (see for example [9]) to the so called
{\it generalized cosine (sine) transform} .  In effect we write
$ e^{ \pm i \, u} = \mp i \partial_u e^{ \pm i \, u} \, $ in (21c)
and then integrate by parts while discarding the surface terms at infinity.
$$ \eqalignno{
{\cal V}'  \, f ({\bf r} ) \,
&= \, - \, {i \over \sqrt{ 2 \pi } } \,
\int_0^\infty \, du \; \Big( \, e^{ - i \, u}  \;
\partial_u [\sqrt{u} \; f ( u \,{\bf r} ) \, ]  \,
+ \, e^{ i \, u}  \;
\partial_u [\sqrt{u} \; f ( - \, u \, {\bf r} ) \, ]  \, \Big)  \cr
&= \, - \, {i \over \sqrt{ 2 \pi } } \,
\int_0^\infty \, du \; \Big( \, { e^{ - i \, u} \over \sqrt{u} } \;
\sqrt{u} \partial_u [\sqrt{u} \; f ( u \,{\bf r} ) \, ]  \,
+ \, { e^{ i \, u} \over \sqrt{u} } \;  \sqrt{u}
\partial_u [\sqrt{u} \; f ( - \, u \, {\bf r} ) \, ]  \, \Big)  \cr
&= \, - \, {i \over \sqrt{ 2 \pi } } \, \sqrt{r} \, \partial_r \,
\Big\{ \sqrt{r} \;
\int_0^\infty \, du \; \Big( \, { e^{ - i \, u} \over \sqrt{u} } \;
f ( u \,{\bf r} )  \,
- \, { e^{ i \, u} \over \sqrt{u} } \; f ( - \, u \, {\bf r} ) \, ] \,
\Big)  \Big\} \, . & (21e) \cr
  }  $$
For the last line we have used the identity
$ \sqrt{u} \, \partial_u  \sqrt{u} \; f ( \pm u \, {\bf r} ) \,
= \pm \, \sqrt{r} \, \partial_r \sqrt{r} \; f ( \pm u \, {\bf r} ) \, .$
The operator $ {\cal V}' \, f ( {\bf r} ) \, $ is equal to
$ {\cal V} \, f ( {\bf r} ) \, $ whenever the latter exists, so from now on
we drop the prime on $ {\cal V}' \, .$

\vskip 0.2in

\noindent{\bf 3. The eigenfunctions
$ {\cal V} \, u_{s{\bf k}} \, $ \hfil\break } \noindent
In this section
we find the eigenfunctions of $ P_s^\lambda =
{\cal V} \, a_s^\lambda \, {\cal V}^{- 1} \,  \, $ which we call
$$ w_{s{\bf k}} \, = {\cal V} \, u_{s{\bf k}} \,   \eqno  (22) $$
and show that these
represent waves which have velocity $ c \, $ along their propagation axis.

The $ {\cal V} \, $ operator sees through any angular variable,
so recalling (15) we need to evaluate \break
$ {\cal V} \, J_{2 s} ( 2 [k r + {\bf k} \cdot {\bf r}]^{1/2} ) \, $
where $ 2 s \, $ is zero or an integer. We will specialize to the case
when $ {\bf k} = ( 0 , 0 , k ) \, $  and the eigenfunction is
$$ u_{s(0,0,k)} ( {\bf r} )
= {1\over 4 \pi} \, e^{2 \, i \, s \, \phi } \,
J_{2 s} \Big( [ 2 k r + 2 k z ]^{1/2} \Big)  \, ,
          \eqno (23) $$
then from (21e) and the integrals (1.13.25), (2.13.27) of [10]
$$ \eqalignno{
w&_{s(0,0,k)} ( {\bf r} )
= \, {\cal V} \, u_{s(0,0,k)} ( {\bf r} ) \cr
&= \, - \, { i \, e^{2 \, i \, s \, \phi } \over 4 \pi \sqrt{ 2 \pi } } \,
\sqrt{r} \, \partial_r \, \Big\{ \sqrt{r} \;
\int_0^\infty \, du \; \Big( \, { e^{ - i \, u} \over \sqrt{u} } \;
J_{2 s} \Big( [ 2 u k r + 2 u k z ]^{1/2} \Big)  \,
- \, { e^{ i \, u} \over \sqrt{u} } \;
J_{2 s} \Big( [ 2 u k r - 2 u k z ]^{1/2} \Big)  \, \Big)  \Big\} \,  \cr
&= \, - \, {e^{2 \, i \, s \, \phi } \over 4 \pi \sqrt{ 2 } } \,
( \sqrt{r} \, \partial_r \,  \sqrt{r} \, ) \;
\Big\{ \, e^{ i ( 1 - 2 s ) \pi  / 4 } \,
\exp \{ i \, ( { { k \, r + k \, z  \over 4 }  }  \} \,
J_{s} ( { k \, r + k \, z  \over 4 }  ) \, \cr
& \qquad \qquad \qquad \qquad \qquad \qquad
- \, e^{ - i ( 1 - 2 s ) \pi  / 4 } \,
\exp \{ - i \, ( { { k \, r - k \, z  \over 4 }  }  \} \,
J_{s} ( { k \, r - k \, z  \over 4 }  ) \, \Big\} .  &  (24a) \cr
  }  $$
Omitting the $ e^{2 \, i \, s \, \phi } \, $ phase factor,
the $ w_{s(0,0,k)} \, $ are everywhere continuous, and are zero at
the origin except for the spin zero case $ s = 0 \, .$
The $ w_{s(0,0,k)} \, $ have similar asymptotic behaviour for various
$ s \, ,$ and are of simpler form when $ s \, $ is half integer.
When for example $ s = 1/2 \, $
$$ \eqalign{
w_{{1\over 2}(0,0,k)} ( {\bf r} )
&= - \, {e^{ i \, \phi } \over 4 \sqrt{\pi}^3 } \, \Big(  [ k r + k z ]^{1/2} \,
\exp ( i [ k r + k z ] / 2 )  \,
- \, [ k r - k z ]^{1/2} \, \exp (  - i [ k r - k z ] / 2 ) \Big) \, \cr
  }  \eqno (25)  $$
which is proportional to $ \exp ( i k z ) \, $ along both halves of the
$ z \, $ axis, so that $ \{ e^{ - i k t } \,  w_{{1\over 2}(0,0,k)} \, \} $
is a unidirectional wave proceeding in the $ + z \, $ direction at velocity
$ c \, .$ For large $ r \, ,$ the $ w_{s(0,0,k)} \, $
are of order $ \sqrt{r} \, ,$ the `extra' factor $ \sqrt{r} \, $
accounted for by the $ 1 / r \, $ inner product space. For reference we
write out the $ w_{s(0,0,k)} \, $ of (24a) in full below
$$ \eqalignno{
w_{s(0,0,k)} ( {\bf r} )
&= { e^{2 \, i \, s \, \phi }  \over 4 \, \pi \, \sqrt{2}} \,
\bigg[ e^{ i {( 1 - 2 s ) \pi \over 4} } \,
\exp \{ i \, k \, \lambda / 2 \} \, \Big\{  ( 2 s - 1 ) \,
J_{s} ( k \, \lambda / 2 ) \, - \, k \, \lambda \,
\Big( J_{s-1} ( k \, \lambda / 2 ) \,
+ \, i \, J_{s} ( k \, \lambda / 2 ) \, \Big) \, \Big\} \, \cr
& \; - \, e^{ - i {( 1 - 2 s ) \pi \over 4} } \,
\exp \{ - i \, k \, \mu / 2 \} \, \Big\{  ( 2 s - 1 ) \,
J_{s} ( k \, \mu / 2 ) \, - \, k \, \mu \,
\Big( J_{s-1} ( k \, \mu / 2 ) \,
- \, i \, J_{s} ( k \, \mu / 2 ) \, \Big) \, \Big\} \, \bigg] \cr
      &  &  (24b)   \cr
  }  $$
where $ \lambda = (r + z ) / 2 \, , \; \mu = (r - z ) / 2 \, .$
A contour plot of these functions reveals the planar nature of the
wave fronts.

\vskip 0.2in

\noindent {\bf 4. The Lorentz operator
$ {\cal V}  \, {\bf K} \, {\cal V}^{- 1} \, $ simplified } \hfil\break
We are interested to see how the Hermitian position operator
$ {\bf r} \, $ transforms under boosts and rotations, so we must evaluate
the commutators $ [ \overline {J}_s^a \, , \, r^b \, ] \, , \;
 [ \overline {K}_s^a \, , \, r^b \, ] \, $ where
$$ \overline {\bf J}_s \, \equiv {\cal V}  \, {\bf J}_s \,
{\cal V}^{- 1} \, , \qquad  \overline {\bf K}_s \,
\equiv {\cal V}  \, {\bf K}_s \, {\cal V}^{- 1} \, .$$
We will first simplify the $ \overline {\bf J}_s \, , \;
\overline {\bf K}_s \, $ operators, and here
we will only consider the $ s = 0 \, $ helicity zero operators
$ \overline {\bf J} \, , \; \overline {\bf K} \, ,$
because the extra helicity components commute with the
position operator.
 
We first need to define the operators $ {\cal U} \, , \;
{\cal U}^{- 1}  \, $ (we also write out again the $ {\cal V} \, , \;
{\cal V}^{- 1}  \, $ of (21) for comparison):
$$ \eqalignno{
{\cal U}  \,
&\equiv \, {1 \over 2 } \, {\cal N} \, {1 \over \sqrt{r}} \,
\big[ {\cal F}_- \,
+ \, {\cal F}_+ \, {\cal P} \, \big] \, \sqrt{r} \, , \qquad \qquad
{\cal U}^{- 1} \, = {1 \over 2 } \, {1 \over \sqrt{r}} \,
\big[ {\cal F}_+ \, + \, {\cal F}_- \, {\cal P} \, \big] \,
\sqrt{r} \, {\cal N} \, , & (26) \cr
{\cal V} \,
&\equiv \, {1 \over 2 } \, {\cal N} \, {1 \over \sqrt{r}} \,
\big[ {\cal F}_- \, - \, {\cal F}_+ \, {\cal P} \, \big] \,
\sqrt{r} \, , \qquad \qquad
{\cal V}^{- 1} \, = {1 \over 2 } \, {1 \over \sqrt{r}} \,
\big[ {\cal F}_+ \, - \, {\cal F}_- \, {\cal P} \, \big] \,
\sqrt{r} \, {\cal N} \, .  \cr
  }  $$
Also
$$ \eqalignno{
&[ {\bf K} \, , ( {\cal N} \, {1 \over \sqrt{r}} \,
{\cal F}_{{}_\pm} \sqrt{r} ) \, ]  = 0 \, ,
\qquad [ {\bf J} \, , ( {\cal N} \, {1 \over \sqrt{r}} \,
{\cal F}_{{}_\pm} \sqrt{r} ) \, ]  = 0 \, , & (27) \cr
&{\bf K}  \, {\cal P} \,
= - {\cal P} \, {\bf K}  \, ,  \qquad \qquad \; \qquad{\bf J} \, {\cal P} \,
= {\cal P} \, {\bf J}  \, & (28) \cr
&{\bf K} \, {\cal V}^{- 1} \,
= {\cal U}^{- 1}  \, {\bf K} \, , \qquad \qquad
{\cal V} \, {\bf K} \,
= {\bf K} \, {\cal U} \,   & (29) \cr
&[ {\bf J} \, , \, {\cal V}\, ]
= [ {\bf J} \, , \, {\cal V}^{- 1} \, ]  = 0 \, . & (30) \cr
      }    $$
The first relation (27) becomes apparent when we write out
$ ( {\cal N} \, {1 \over \sqrt{r}} \,
{\cal F}_{{}_\pm} \sqrt{r} ) \, \psi ( {\bf r} ) \, $ in the form
 $$ \eqalignno{
( {\cal N} \, {1 \over \sqrt{r}} \, {\cal F}_{{}_\pm} \sqrt{r} ) \,
\psi ( {\bf r} ) \,
&\equiv \sqrt{ 2 \over \pi } \, \int_0^\infty \, du  \;
e^{ \pm i \, u} \; \sqrt{u} \; \psi ( u \, {\bf r}) \, , \cr
  }  $$
then one can see that $ {\bf K} , {\bf J} \, $ only act on the $ {\bf r} \, $
in the argument of $ \psi \, .$
And (29) follows from (27), (28).
The relation (30) means that
$$ \overline {\bf J} \, \equiv {\cal V} \, {\bf J} \, {\cal V}^{- 1} \,
= \, {\bf J} \, . $$
Following from (29)
$$ \eqalign{
&\overline {\bf K} \, \equiv {\cal V} \, {\bf K} \, {\cal V}^{ \dag } \,
= \, ( {\cal V} \, {\cal U}^{ \dag } ) \, {\bf K} \,
= \, {\bf K} \, ( {\cal U} \, {\cal V}^{ \dag } ) \, .
 \cr           }   \eqno (31) $$

We can simplify the $ ( {\cal V} \, {\cal U}^{ \dag } ) \, ,
\; ( {\cal U} \, {\cal V}^{ \dag } ) \, $ unitary operators of (31) above
which are adjoints of each other.
We need the further identities (referred to in the appendix)
$$ \eqalign{
&{\cal F}_+ \, {\cal F}_+
= \, - \, i \, ( {\cal H}_e - {\cal H}_o ) \, ,
\qquad \qquad {\cal F}_- \, {\cal F}_-
= \, i \, ( {\cal H}_e - {\cal H}_o ) \, ,   \cr
&{\cal F}_+ \, {\cal F}_-
= 2 \, - \, i \,( {\cal H}_e + {\cal H}_o )  \, ,
\qquad \quad {\cal F}_- \, {\cal F}_+
= 2 \, + \, i \, ( {\cal H}_e + {\cal H}_o )  \,   \cr
   }   \eqno (32)     $$
where $ {\cal H}_e , \, {\cal H}_o \, $ are the Hilbert transforms of
even, odd functions:
$$ \eqalign{
{\cal H}_e \, f(r , \theta , \phi ) &\equiv \, - \, { 2 r \over \pi } \,
\int_0^\infty { f(t , \theta , \phi ) \over r^2 - t^2 } \, dt \, , \qquad \quad
{\cal H}_e \, f({\bf r}) = \, - \, { 2  \over \pi } \,
\int_0^\infty { \, f( u {\bf r} ) \over 1 - u^2 } \,
du \, , \cr
{\cal H}_o \, f(r , \theta , \phi ) &\equiv  \, - \,  { 2 \over \pi } \,
\int_0^\infty { t \, f(t , \theta , \phi ) \over r^2 - t^2 } \, dt \, ,
\qquad \quad
{\cal H}_o \, f({\bf r}) =  \, - \,  { 2 \over \pi } \,
\int_0^\infty { u \, f( u {\bf r} ) \over 1 - u^2 } \,
du \, , \cr }
  \eqno (33)   $$
and also we note that
$$ {\cal N} \, ( {1 \over \sqrt{r} } \, {\cal H}_e   \sqrt{r} ) \,
{\cal N} =  ( {1 \over \sqrt{r} } \, {\cal H}_o   \sqrt{r} ) \, , \qquad
\, {\cal N} \, ( {1 \over \sqrt{r} } \, {\cal H}_o   \sqrt{r} ) \,
{\cal N} =  ( {1 \over \sqrt{r} } \, {\cal H}_e   \sqrt{r} ) \, .
   \eqno  (34)   $$
Then with the aid of (32), (34)
$$ \eqalignno{
( {\cal V} \, {\cal U}^{ \dag } ) \,
&= {1 \over 4 } \, {\cal N} \, {1 \over \sqrt{r}} \,
\big[ {\cal F}_- \, - \, {\cal F}_+ \, {\cal P} \, \big] \,
\big[ {\cal F}_+ \,
+ \, {\cal F}_- \, {\cal P} \, \big] \, \sqrt{r} \, {\cal N} \,  \cr
&= {1 \over 2 } \, {\cal N} \, {1 \over \sqrt{r}} \,
\big[  i \, ( {\cal H}_e + {\cal H}_o )  \,
+ \, i \, ( {\cal H}_e - {\cal H}_o ) \, {\cal P} \, \big] \,
 \sqrt{r} \, {\cal N} \,  \cr
&=  {1 \over 2 } \, i \, {1 \over \sqrt{r}} \,
\big[  ( {\cal H}_e + {\cal H}_o )  \,
- \, ( {\cal H}_e - {\cal H}_o ) \, {\cal P} \, \big] \, \sqrt{r} \,
\equiv  i \, {1 \over \sqrt{r}} \, {\cal G}_-  \, \sqrt{r} \, &  (35) \cr
\noalign{\noindent and similarly }
( {\cal U} \, {\cal V}^{ \dag } ) \,
&=  {1 \over 2 } \, i \, {1 \over \sqrt{r}} \,
\big[  ( {\cal H}_e + {\cal H}_o )  \,
+ \, ( {\cal H}_e - {\cal H}_o ) \, {\cal P} \, \big] \,
 \sqrt{r} \,
\equiv  i \, {1 \over \sqrt{r}} \, {\cal G}_+  \, \sqrt{r} \, &  (36) \cr
\noalign{\noindent where }
{\cal G}_{{}_\pm}
&\equiv \, {1 \over 2} \, \left[ \, ( {\cal H}_e + {\cal H}_o )  \,
\pm \,  ( {\cal H}_e - {\cal H}_o )  \, {\cal P} \,  \right] \,  & (37a)   \cr
{\cal G}_{{}_\pm}  f({\bf r})
&= \,  { 1 \over \pi } \, \int_0^\infty
\Big( \,  {  f( u {\bf r}) \over u - 1 } \,
\mp \, {  f( - u {\bf r}) \over u + 1  } \, \Big) \;
du \, . & (37b) \cr
     }  $$
Finally substituting (35), (36) into (31) we have the transformed boost
operator
$$ \eqalignno{
\overline {\bf K} \, &\equiv  \, {\cal V} \, {\bf K} \, {\cal V}^{ \dag } \,
= \, {\bf K}  \, ( i \, {1 \over \sqrt{r} } \,
{\cal G}_+  \sqrt{r} \, ) \,
=  \, (  i \, {1 \over \sqrt{r} } \, {\cal G}_-   \sqrt{r} \, ) \,
{\bf K}  \,  & (38)  \cr
      }    $$
where $ {\bf K} \equiv \, (  - \, i \, r \, {\bx \nabla} ) \, .$
Note that $ \overline {\bf K} \, $ is not a local (differential) operator
like $ {\bf K} \, .$

\vskip 0.2in

\noindent{\bf 5. The position 4-vector operator \hfil\break }
\noindent The simplification (38) allows us to to calculate the commutator
$ [ \overline {K}^a \, , \, r^b \, ] \, .$
For the position operator $ {\bf r} \, $ to be the space
part of a 4-vector $ r^\lambda \, \equiv ( r^0 , {\bf r} ) \, ,$
then $ r^0 \, $ must satisfy both of the following:
$$ [ \overline {K}^a \, , \, r^b \, ] \, = i \, \delta^{a b} r^0 \, , \qquad
 [ \overline{\bf K} \, , \, r^0 \, ] \, = i \, {\bf r} \, . \eqno  (39) $$

To calculate $ [ \overline {K}^a \, , \, r^b \, ] \, $ we must
first evaluate $ [ {\cal G}_\pm \, , \, {\bf r} ] \, .$
Recalling (37b) then
$$ \eqalignno{
{\cal G}_\pm \, \big( \, {\bf r} \, f({\bf r}) \, \big)
&= \,  { 1 \over \pi } \, \int_0^\infty
\Big( \,  {  u {\bf r} \, f( u {\bf r}) \over u - 1 } \,
\pm \, { u {\bf r} \,  f( - u {\bf r}) \over u + 1 } \,
\Big) \; du \,     \cr
&= \,  { 1 \over \pi } \, {\bf r} \, \int_0^\infty
\Big( \,  {  ( u - 1 + 1 ) \, f( u {\bf r}) \over u - 1 } \,
\pm \, { ( u + 1 - 1 ) \,  f( - u {\bf r}) \over u + 1 } \,
\Big) \; du \,    \cr
&= \, {\bf r} \, {\cal G}_\pm  f({\bf r}) \,
+ \, { 1 \over \pi } \, {\bf r} \, \int_0^\infty
( \,  f( u {\bf r}) \,
\pm \,  f( - u {\bf r}) \, ) \; du \, ,  & (40)   \cr
     }  $$
and multiplying (37b) from the left by $ {\bf r} \, $ and then subtracting
from (40) yields the operator identity
$$ \eqalignno{
&[ \, {\cal G}_\pm \, , \, {\bf r} \, ] \, = \, {\bf r} \,
{\cal Z}_\pm  \, ,  &  (41) \cr
\noalign{\noindent where }
&{\cal Z}_\pm f({\bf r}) \, \equiv \, { 1 \over \pi } \,
\int_0^\infty ( \,  f( u {\bf r}) \,
\pm \,  f( - u {\bf r}) \, ) \; du \, .  &  (42)  \cr
  }    $$
Also with similar methods to the above we find
$$ \eqalignno{
{\cal G}_\pm \,  \big( \, r \, f({\bf r}) \, \big)
&= \, r \, {\cal G}_\mp  f({\bf r}) \,
+ \, { 1 \over \pi } \, r \, \int_0^\infty ( \,  f( u {\bf r}) \,
\mp \,  f( - u {\bf r}) \, ) \; du \,    \cr
\noalign{\noindent or }
{\cal G}_\pm \,  r
&= \, r \, {\cal G}_\mp \, + \, r \, {\cal Z}_\mp  \, .  & (43)   \cr
   }    $$
With the commutator (41) we can now evaluate
$ [ \overline {K}^a \, , \, r^b \, ] \, :$
$$ \eqalignno{
[ \, \overline {K}^a \, , \, r^b \, ] \,
&= [ \, {K}^a  \, (  i \, {1 \over \sqrt{r} } \,
{\cal G}_+  \sqrt{r} \, )  \, , \, r^b \, ] \,  \cr
&= [ \, {K}^a  \,  \, , \, r^b \, ] \, (  i \, {1 \over \sqrt{r} } \,
{\cal G}_+  \sqrt{r} \, ) \,
+ \, {K}^a  \, ( i \, {1 \over \sqrt{r} } \,
[ \, {\cal G}_+ \, , \, r^b \, ]  \sqrt{r} \, ) \,   \cr
&=  \, - i \, \, \delta^{ab} \, r \, (  i \, {1 \over \sqrt{r} } \,
{\cal G}_+  \sqrt{r} \, ) \,
+ \, {K}^a  \, ( \, {1 \over \sqrt{r} } \, r^b \,
{\cal Z}_+ \, \sqrt{r} \, ) \, \rightarrow  \, - i \, \, \delta^{ab} \, r \,
( i \, {1 \over \sqrt{r} } \, {\cal G}_+  \sqrt{r} \, ) \, &  (44) \cr
  }    $$
where for the last line we put $ {\cal Z}_+ \, \sqrt{r} \, = 0 \, .$
This is a boundary condition on the wavefunction requiring that
$$ {\cal Z}_+ \big( \sqrt{r} \psi ({\bf r}) \, \big) \,
\equiv \, { 1 \over \pi } \, \sqrt{ r } \,
\int_0^\infty du \; \sqrt{ u } \, [  \psi ( u {\bf r}) \,
+ \,  \psi ( - u {\bf r}) \, ] \; = 0 \, . \eqno (45) $$
The above is automatically satisfied if $ \psi ({\bf r}) \, $ is odd under
the parity transformation. When $ \psi ({\bf r}) \, $ is even under
parity then (45) is equivalent to $ \int_0^\infty dr \;
\sqrt{ r } \,  \psi ( r , \theta , \phi ) \, = 0 \, .$

Recalling (39), the formula (44) suggests that
$$ \eqalignno{
r^0 &= \, - \, r \, ( i \, {1 \over \sqrt{r} } \,
{\cal G}_+ \sqrt{r} \, ) \, \cr
&= - \, ( i \, {1 \over \sqrt{r} } \,
{\cal G}_-  \sqrt{r} \, ) \, r \, + \, i \, r ( i \, {1 \over \sqrt{r} } \,
{\cal Z}_+  \sqrt{r} \, ) \, = - \, ( i \, {1 \over \sqrt{r} } \,
{\cal G}_-  \sqrt{r} \, ) \, r \, \cr
    }   $$
where the second line results from (43) and (45). So we can write
two equivalent forms for $ r^0 \, :$
\setbox4=\vbox{\hsize 19pc \noindent \strut \hfil
$ \displaystyle r^0 = \, - \, r \, ( i \, {1 \over \sqrt{r} } \,
{\cal G}_+ \sqrt{r} \, ) \,
= \, - \, ( i \, {1 \over \sqrt{r} } \, {\cal G}_-  \sqrt{r} \, ) \, r \, .$
\strut}
$$ \boxit{ \box4 }   \eqno (46)  $$
We must still check the commutator
$ [ \overline {\bf K} \, , \, r^0 \, ] \, ,$ recalling from (38)
and (46) that both the operators $ \overline {\bf K} \, , \; r^0 \, $
can be written in either of two ways:
$$ \eqalignno{
[ \overline {\bf K} \, , \, r^0 \, ]
&= \overline {\bf K} \, r^0 \, - \, r^0 \, \overline {\bf K} \, \cr
&= \, - \, {\bf K}  \, ( i \, {1 \over \sqrt{r} } \,
{\cal G}_+  \sqrt{r} \, ) \,  ( i \, {1 \over \sqrt{r} } \,
{\cal G}_-  \sqrt{r} \, ) \, r \,
+ \, r \, (  i \, {1 \over \sqrt{r} } \, {\cal G}_+   \sqrt{r} \, ) \,
 \, ( i \, {1 \over \sqrt{r} } \,
{\cal G}_-  \sqrt{r} \, )  {\bf K} \,  \cr
&= \, - \, {\bf K}  \, r \, + \, r \, {\bf K} \,  = \, i \, {\bf r} \,
   &  (47)  \cr
  }    $$
which is the required result, and we have used the identity
$$ ( {i \over \sqrt{r} } \, {\cal G}_-   \sqrt{r}  ) \,
( {i \over \sqrt{r} } \, {\cal G}_+   \sqrt{r} ) \,
= \, ( {\cal V} \, {\cal U}^{ \dag } ) \,
( {\cal U} \, {\cal V}^{ \dag } ) \, = 1  \eqno  (48) $$
recalling (35), (36).
The results (44), (47) show the 4-vector property of
$$ r^\lambda = \Big( \, - \, r \, ( i \, {1 \over \sqrt{r} } \, {\cal G}_+
\sqrt{r} \, ) \, , \; {\bf r} \, \Big) \,   $$
only assuming (45). Furthermore we see from (46) that
$$ \eqalignno{
( r^0 )^2 &= \, r \, ( i \, {1 \over \sqrt{r} } \, {\cal G}_+ \sqrt{r} \, ) \,
( i \, {1 \over \sqrt{r} } \,
{\cal G}_-  \sqrt{r} \, ) \, r \, = r^2 \, &  (49)  \cr
    }   $$
so that $ r^\lambda \, $ is a null 4-vector. Finally the
$ r^0 \, , \; {\bf r} \, $ components of $ r^\lambda \, $ commute as
$$ \eqalignno{
[ \, r^0 \, , \, {\bf r} \, ] \,
&=  \, - \, r   \, (  i \, {1 \over \sqrt{r} } \,
[ {\cal G}_+   \, , \, {\bf r} \, ] \, \sqrt{r} \, )
=  \, - \, r   \, (  i \, {1 \over \sqrt{r} } \,
{\bf r} \, {\cal Z}_+  \, \sqrt{r} \, ) \, = \, 0  \cr
  }    $$
with (45).

\vskip 0.2in

\noindent{\bf 5. Discussion \hfil\break } \noindent
We have achieved our aims (A) to (E) of the Introduction. And
we have shown that the position operator $ {\bf r} \, $
has covariant meaning, in that it is the space part of a null 4-vector
(as discussed in the last section, this 4-vector property of
$ r^\lambda \, $ requires the wavefunction to be subject to the boundary
condition (45)). The difficulties with the position operator
have been well known since the inception
of relativistic quantum mechanics: the eigenfunctions of
the Newton-Wigner [3] operator $ ( {\bf r}_{NW} ) \, $ are smeared out in
space, which renders problematic the exact meaning of $ {\bf r} \, $
when, for example, potential terms are included in the Hamiltonian.
From our viewpoint
the difficulties in the usual theory arise from the inner product (3),
which is a direct consequence of the momentum operator
being $ {\bf p} = ( - i {\bx \nabla} ) \, .$ Then the inner product (3)
disallows the `natural' position operator $ {\bf r} \, .$

Our approach has been to allow other possibilities for the momentum
operator, only requiring that it is the space part of a 4-vector and that
the eigenfunctions of the energy-momentum operator
have plane wave character. Because the momentum operators presented
here are Hermitian with respect to the simple $ 1 / r \, $ inner product
space of (7), this in turn allows for the natural position
operator $ {\bf r} \, .$ The `cost' we have to pay for this transparency of
the position operator is that the momentum operators
as well as their eigenfunctions are more complicated.

\beginsection Appendix

\noindent{\it The angular phase factor
$ e^{2 \, i \, s \, f ( {\bf \hat r} , {\bf \hat k} ) } \, $ } \hfil\break
In spherical coordinates
$$ \eqalignno{
{\bf \hat r} &= ( \cos \phi \, \sin \theta , \, \sin \phi \, \sin \theta , \,
 \cos \theta ) \cr
{\bf \hat k} &= ( \cos \phi_k \, \sin \theta_k , \,
\sin \phi_k \, \sin \theta_k , \, \cos \theta_k ) \cr
\noalign{\noindent then we give Derrick's formula ((39) of [8]) }
e^{ i \, f ( {\bf \hat r} , {\bf \hat k} ) } \,
&= [ 2 / (1 + {\bf \hat k} \cdot {\bf \hat r} ) ]^{1/2} \,
( e^{ i \phi } \, \cos {\scriptstyle {1 \over 2}} \theta \,
\cos {\scriptstyle {1 \over 2}} \theta_k \, + \,
e^{ i \phi_k } \, \sin {\scriptstyle {1 \over 2}} \theta \,
\sin {\scriptstyle {1 \over 2}} \theta_k \, ) & (A1) \cr
\noalign{\noindent and in the particular case when
$ {\bf \hat k} = ( 0 , 0 , 1) $ so that $ \theta_k = 0 \, ,$ then  }
e^{ i \, f ( {\bf \hat r} , {\bf \hat k} ) } \,
&\rightarrow [ 2 / (1 + {\hat r}_3 ) ]^{1/2} \,
( e^{ i \phi } \, \cos {\scriptstyle {1 \over 2}} \theta \, )
= e^{ i \phi } \, & (A2)  \cr
   }  $$

\vskip 0.1 in

\noindent{\it The unitary operators $ {\cal V} \, $
 \hfil\break }
\noindent The unitary operators $ {\cal V} \, $ of (21) can be constructed
from the
parity $ {\cal P} \, ,$ inversion $ {\cal N} \, ,$ and Fourier
transform operators $ {\cal F} \, :$
$$  \eqalignno{
&{\cal P} \, f(r , \theta , \phi )
\equiv  f (r , \pi - \theta , \phi + \pi) \, &  (A3)  \cr
&{\cal N} \, f(r , \theta , \phi ) \equiv { 1 \over r^2 } \,
f ({ 1 \over r } \,  , \theta , \phi ) \, &  (A4)  \cr
&{\cal F}_{c \atop s } \, f(r , \theta , \phi )
\equiv \, \sqrt{2 \over \pi } \, \int_0^\infty \,
{\cos (  r \, t ) \atop \sin (  r \, t ) }  \;
f (t , \theta , \phi ) \, dt  \, &  (A5) \cr
&{\cal F}_{{}_\pm} \,
\equiv ( {\cal F}_c \, \pm \, i \, {\cal F}_s ) \, &  (A6)  \cr
\noalign{\noindent with the adjoint properties   }
&{\cal P}^{\dag} =  \, {\cal P} \, , \qquad
{\cal N}^{\dag} =  \, {\cal N} \, , \qquad
( {1 \over \sqrt{r}} \, {\cal F}_{c \atop s } \sqrt{r} )^{\dag}  \,
= \, ( {1 \over \sqrt{r}} \, {\cal F}_{c \atop s } \sqrt{r} ) \, . & (A7) \cr
\noalign{\noindent
(The self-adjoint property $  {\cal N}^{\dag} =  \, {\cal N} \, $
can be shown by a change of variables
$ r \rightarrow 1 / u \, $ within the scalar product $ {\cal L}_{1/r} \, ,$
and the adjoint property of
$ ( {1 \over \sqrt{r}} \, {\cal F}_{c \atop s } \sqrt{r} ) \, $
follows by changing the order of integration within the scalar product.)
Also }
&{\cal P} \, {\cal P} = {\cal N} \, {\cal N}
= {\cal F}_c \, {\cal F}_c \,
= {\cal F}_s \, {\cal F}_s \, = \, 1 \, .  &  (A8) \cr
  }    $$
For the latter property
$ {\cal F}_c \, \big( {\cal F}_c f(r) \big) = \,
{\cal F}_s \, \big( {\cal F}_s f(r) \big) = \,
f(r) \, $ to hold always, the definitions of
$ {\cal F}_c \, {\cal F}_s \, $ must be extended
[9] to the so called
{\it generalized cosine (sine) transform}
when the integral of (A5) does not exist. Then
$$ {\cal F}_c \rightarrow - \, {1 \over r} \, {\cal F}_{s} \, \partial_r
\, , \qquad
{\cal F}_s \rightarrow  {1 \over r} \, {\cal F}_{c} \, \partial_r
\, , \qquad  {\cal F}_{{}_\pm} \, \rightarrow \pm \, {i \over r} \,
{\cal F}_{{}_\pm} \, \partial_r  \,  \eqno (A9) $$
which is effectively an integration by parts within the integral (A5)
while discarding the surface terms.
When the conventional cosine (sine) transform of (A5) does exist, it agrees
with the extended definition (A9). The extension procedure of (A9) can be
repeated.
Then the operators $ ( {1 \over \sqrt{r}} \, {\cal F}_c \sqrt{r} ) , \,
( {1 \over \sqrt{r}} \, {\cal F}_s \sqrt{r} ) \, $
as well as $ {\cal P} , \, {\cal N} , \, $  are unitary.

The operators $ {\cal F}_c \, , {\cal F}_s \, $
do not commute but combine as follows [11]:
$$ {\cal F}_s {\cal F}_c \, =  - \, {\cal H}_e \, , \qquad
{\cal F}_c {\cal F}_s \, = \, {\cal H}_o \, \eqno (A10) $$
where $ {\cal H}_e , \, {\cal H}_o \, $ are the Hilbert transforms of
even, odd functions:
$$ \eqalign{
{\cal H}_e \, f(r , \theta , \phi ) &\equiv \, - \, { 2 r \over \pi } \,
\int_0^\infty { f(t , \theta , \phi ) \over r^2 - t^2 } \, dt \, , \qquad \quad
{\cal H}_e \, f({\bf r}) = \, - \, { 2  \over \pi } \,
\int_0^\infty { \, f( \lambda {\bf r} ) \over 1 - \lambda^2 } \,
d\lambda \, , \cr
{\cal H}_o \, f(r , \theta , \phi ) &\equiv  \, - \,  { 2 \over \pi } \,
\int_0^\infty { t \, f(t , \theta , \phi ) \over r^2 - t^2 } \, dt \, ,
\qquad \quad
{\cal H}_o \, f({\bf r}) =  \, - \,  { 2 \over \pi } \,
\int_0^\infty { \lambda \, f( \lambda {\bf r} ) \over 1 - \lambda^2 } \,
d\lambda \, . \cr }
  \eqno (A11)   $$
Then it follows that
$$ \eqalign{
&{\cal F}_+ \, {\cal F}_+
= \, - \, i \, ( {\cal H}_e - {\cal H}_o ) \, ,
\qquad \qquad {\cal F}_- \, {\cal F}_-
= \, i \, ( {\cal H}_e - {\cal H}_o ) \, , \qquad  \cr
&{\cal F}_+ \, {\cal F}_-
= 2 \, - \, i \,( {\cal H}_e + {\cal H}_o )  \, ,
\qquad \quad {\cal F}_- \, {\cal F}_+
= 2 \, + \, i \, ( {\cal H}_e + {\cal H}_o )  \, . \qquad     \cr
   }   \eqno (A12)     $$
The identities (A12) allow us to construct
the unitary operator $ {\cal V} \, $ and its adjoint:
$$ \eqalignno{
{\cal V} \,
&\equiv \, {1 \over 2 } \, {\cal N} \, {1 \over \sqrt{r}} \,
\big[ {\cal F}_{{}_-} \,
- \, {\cal F}_{{}_+} \, {\cal P} \, \big] \, \sqrt{r} \, , \qquad \qquad
{\cal V}^{- 1} \, = {1 \over 2 } \, {1 \over \sqrt{r}} \,
\big[ {\cal F}_{{}_+} \, - \, {\cal F}_{{}_-} \, {\cal P} \, \big] \,
\sqrt{r} \, {\cal N} \, , & (A13) \cr
  }  $$
then the necessary properties
$ {\cal V} \, {\cal V}^{- 1} \, = \, 1 \,  $ etc
follow from (A8), as the parity operator $ {\cal P} \, $ commutes with
$ {\cal F}_{{}_\pm} \, $ and $ {\cal N} \, ,$ also $ {\cal P}^2
= {\cal N}^2 =1 \, .$
Writing out the operator $ {\cal V} \, f ( r , \theta , \phi ) \, :$
$$ \eqalignno{
{\cal V}  \, f ( r , \theta , \phi ) \,
&\equiv {1 \over 2 } \,{\cal N} \, \big[
( {1 \over \sqrt{r}} {\cal F}_- \sqrt{r} ) \,
- \, ( {1 \over \sqrt{r}} {\cal F}_+ \sqrt{r} ) \,
{\cal P} \, \big] \, f ( r , \theta , \phi ) \,  & (A14a)  \cr
&= {1 \over \sqrt{ 2 \pi } } \, {1 \over \sqrt{r}^3 }
\int_0^\infty \, dt  \; \big[ e^{ - i \, t / r } \; \sqrt{t} \;
f ( t , \theta , \phi ) \, - \, e^{ i \, t / r } \; \sqrt{t} \;
f ( t , \pi - \theta , \phi + \pi ) \, \big]  & (A14b) \cr
&= {1 \over \sqrt{ 2 \pi } } \,
\int_0^\infty \, du  \; \big[ e^{ - i \, u } \; \sqrt{u} \;
f ( u r , \theta , \phi ) \, - \, e^{ i \, u } \; \sqrt{u} \;
f ( u r , \pi - \theta , \phi + \pi ) \, \big]  & (A14c) \cr
\noalign{\noindent \baselineskip16pt where we have substituted
$ t = u \, r \, .$ Then changing to cartesian coordinates we can write }
{\cal V}  \, f ( {\bf r} ) \,
&= {1 \over \sqrt{ 2 \pi } } \,
\int_{ - \infty }^\infty \, du \; \, e^{ - i \, u} \; \sqrt{|u|} \;
\hbox{sgn} (u) \;  f ( u \, {\bf r}) \,  & (A14d) \cr
\noalign{\noindent \baselineskip16pt
which is (21d). Then with the operator substitutions of (A9) }
{\cal V}  \, f ({\bf r} ) \,
&\rightarrow \, - \, {i \over \sqrt{ 2 \pi } } \,
\int_0^\infty \, du \; \Big( \, e^{ - i \, u}  \;
\partial_u [\sqrt{u} \; f ( u \,{\bf r} ) \, ]  \,
+ \, e^{ i \, u}  \;
\partial_u [\sqrt{u} \; f ( - \, u \, {\bf r} ) \, ]  \, \Big)  \cr
  }  $$
which is (21e).

\beginsection References

\frenchspacing \baselineskip 14 pt
\noindent \item{[1]} Foldy L.L., Phys. Rev. {\bf 102}, 568-561 (1956)
\noindent \item{[2]} Schweber S.S., {\it An Introduction to Relativistic
Quantum Field Theory}  (Harper and Row, New York, 1961) p57
\noindent \item{[3]} Newton T. D. and Wigner E. P.
Rev. Mod. Phys. {\bf 21}, 400-406 (1949)
\noindent \item{[4]} Derrick G. H., J. Math. Phys. {\bf 28 },
1327-1340 (1987)
\noindent \item{[5]} Barut A. O. and Ramsmussen W., J. Phys. B {\bf 6},
1695-1712 (1973)
\noindent \item{[6]} Peres A., J. Math. Phys. {\bf 9}, 785-789 (1967)
\noindent \item{[7]} Kastrup H. A., Phys. Rev. B {\bf 140}, 183-186 (1965)
\noindent \item{[8]} Derrick G. H., J. Math. Phys. {\bf 29 },
636-641 (1988)
\noindent \item{[9]} Zemanian A. H., {\it Generalized Integral
Transformations}  (Interscience Publishers, New York, 1968) p163
\noindent \item{[10]} Bateman H., {\it Tables of Integral Transforms,
Vol 1} (McGraw-Hill, New York, 1954) Formulae (1.13.25) and (2.13.27)
\noindent \item{[11]} Rooney P. G., Can. J. Math. {\bf 24},
1198-1216 (1972) Sec.8

\vfill\eject
\nonfrenchspacing

\bye

\end